\documentclass[a4paper]{jpconf} 
\usepackage{graphicx}
\usepackage{xspace}
\usepackage{multicol}

\newcommand{\ape}{APE\xspace}
\newcommand{\apenet}{APEnet\xspace}
\newcommand{\apenetp}{APEnet+\xspace}
\newcommand{\apelink}{APElink\xspace}
\newcommand{\apelinkp}{APElink+\xspace}
\newcommand{\ie}{\textit{i.e.}\xspace}
\newcommand{\eg}{\textit{e.g.}\xspace}
\newcommand{\etc}{\textit{etc.}\xspace}
\newcommand{\quong}{QUOnG\xspace}

\begin{document}
\title{APEnet+: high bandwidth 3D torus direct network for petaflops scale commodity clusters}

\author{R Ammendola$^{1}$, A Biagioni$^2$, O Frezza$^2$, F Lo Cicero$^2$, A Lonardo$^2$,\\ P S Paolucci$^2$, D Rossetti$^2$, A Salamon$^1$, G Salina$^1$, F Simula$^3$,\\ L Tosoratto$^2$ and P Vicini$^2$}


\address{$^1$ INFN Tor Vergata, Roma}
\address{$^2$ INFN Roma, Roma}
\address{$^3$ Sapienza Universit\`{a} di Roma, Roma}

\ead{francesca.locicero@roma1.uniroma1.it, piero.vicini@roma1.infn.it}

\begin{abstract} 

  
%
We describe herein the \apelinkp board, a PCIe interconnect adapter
featuring the latest advances in wire speed and interface technology
plus hardware support for a RDMA programming model and experimental
acceleration of GPU networking; this design allows us to build a low
latency, high bandwidth PC cluster, the \apenetp network, the new
generation of our cost-effective, tens-of-thousands-scalable cluster
network architecture.

Some test results and characterization of data transmission of a
complete testbench, based on a commercial development card mounting an
Altera\textsuperscript{\textregistered} FPGA, are provided.

\end{abstract}

\section{Introduction}
The Array Processor Experiment (\ape) is a custom design for HPC
targeting the field of Lattice QCD, started by the Istituto Nazionale
di Fisica Nucleare and partnered by a number of physics institutions
all over the world, that since its start in 1984 has developed four
generations of custom machines
~\cite{APE,apeNEXT-lat1,apeNEXT-lat2,apeNEXT}. Leveraging on the
acquired know-how in networking and re-employing the gained insights,
a spin-off project called \apenet~\cite{apenet2004,apenet2005}
developed an interconnect board that allows assembling a PC cluster \`a
la \ape with off-the-shelf components.

Following further developments funded by EU projects (FP 6
SHAPES~\cite{SHAPES1,DNP1} and FP 7 EURETILE), the \apenet project
evolved into \apenetp~\cite{ape3}; its achievement is the design of
the \apelinkp host adapter, which integrates both a network interface
and a switching component, bringing in state-of-the-art wire speeds
for the links and a PCIe X8 gen2 host interface. With this latest push
to higher bandwidth, low power and low cost of the data transmission
system, we are encompassing not only a broader range of intensive
numerical algorithms (Lattice QCD is our primary but not exclusive
concern), but also the field of acquisition systems for modern
particle and astroparticle experiments (sLHC, ILC, CLIC, NA62\ldots).



The outlook of this article is as follows: the first section explains
the global network architecture; the second one gives the details of
the host board; the third one outlines the software stack provided by
the programming environment; the fourth one sketches the current
deployment of \apenetp hardware in the framework of our \quong HPC
initiative; the fifth and final one gives conclusions and outlines to
future work.

\section{The \apenetp hardware} 
\label{sec:hw}
%
%
The \apenetp interconnect is our low latency, high bandwidth
packet-based direct network, supporting state-of-the-art link wire
speeds and a PCIe X8 gen2 host connection. On this network, the
computing host --- \eg a multi-core CPU optionally paired with GPU ---
is equipped with one \apelinkp board and made into a node of the
cluster. The nodes are connected by point-to-point links to form a 3D
torus in a cubic mesh; each node communicates with each of its 6
neighbours along the $X+$, $X-$, $Y+$, $Y-$, $Z+$ and $Z-$ directions
by bi-directional full-duplex communication channels.

Size envelope (header+footer) of packets is hard-coded and fixed, while
payload size is variable; packets are auto-routed to their final
destinations according to wormhole dimension-ordered static routing,
with the system taking care of dead-lock avoidance.

\begin{figure}[h]
\includegraphics[width=.4\textwidth]{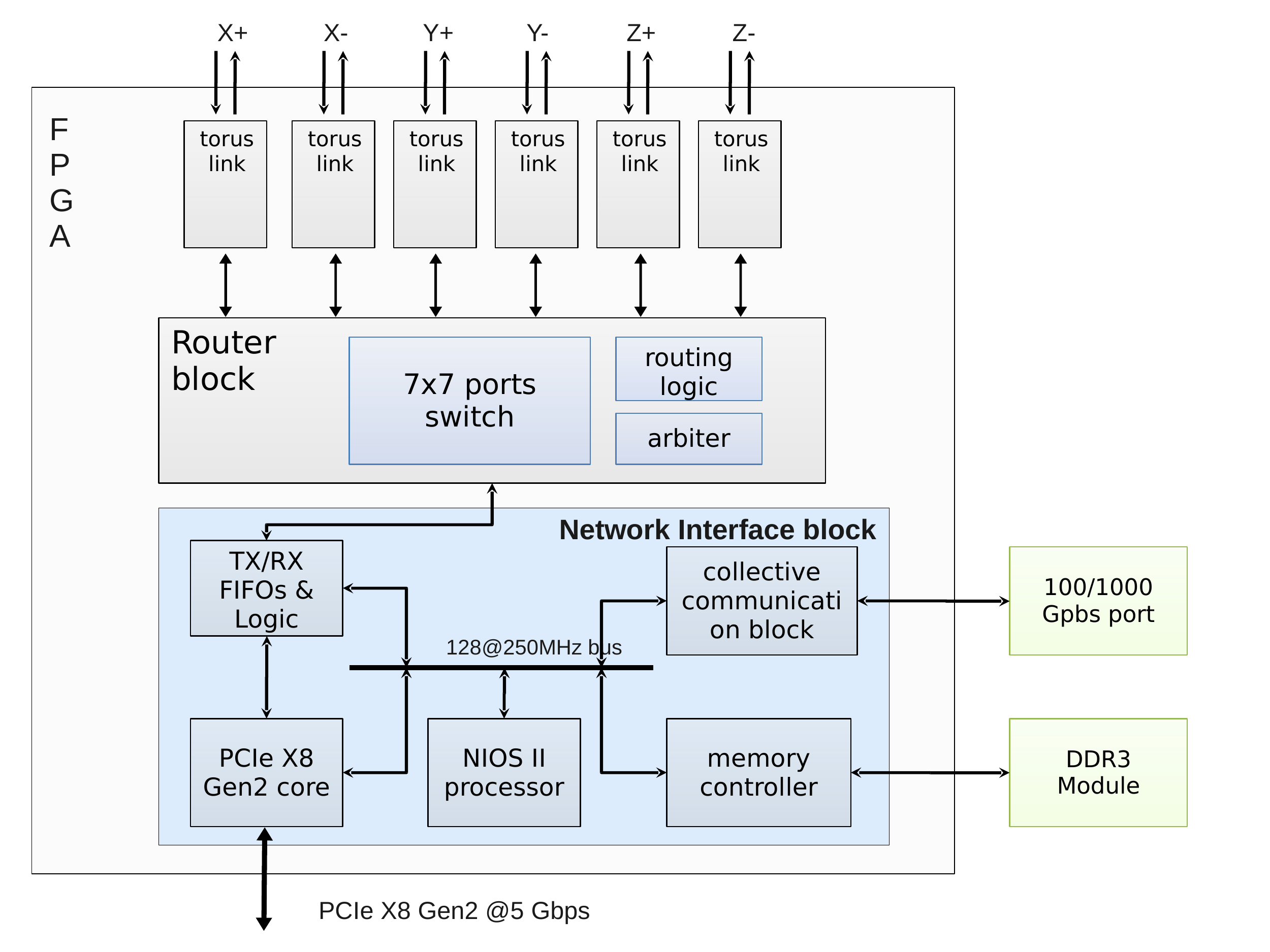}\hspace{2pc}%
\begin{minipage}[b]{14pc}\caption{\label{fig:internals}Internal FPGA block architecture.}
\end{minipage}
\end{figure}

The hardware block structure, depicted in Figure~\ref{fig:internals}, is
split into a so called \emph{network interface} --- the packet
injection and processing logic comprising PCIe, TX/RX logic, \etc ---
a \emph{router} component and multiple \emph{torus links}.

The \apelinkp \emph{network interface} has basically two main tasks:
\begin{itemize}
\item On the transmit data path, it gathers data coming in from the
  PCIexpress port, fragmenting the data stream into packets which are
  forwarded to the relevant destination ports, depending on the
  requested operation.
\item On the receive side, it provides hardware support for the RDMA
  programming model, implementing the basic RDMA capabilities (PUT and
  GET semantics) at the firmware level.
\end{itemize}
Within this block, the addition of a NIOS II 32 bit embedded
micro-controller --- a standard
Altera\textsuperscript{\textregistered} Intellectual Property ---
simplifies some tasks along the path of the received packets.

The \emph{routing block} takes care of examining the packet header and
resolving the destination address into a proper path across the switch
according to the chosen routing algorithm.

The \emph{torus link block} manages the data flow by encapsulating the
\apenetp packets into a light, low-level, \emph{word-stuffing}
protocol able to detect transmission errors via CRC. It implements two
virtual channels~\cite{Dally} and proper flow-control logic on each RX
link block to guarantee deadlock-free operations.
                               

\section{The \apelinkp card}
\label{apelink_card}

For the design of the building brick of the \apenetp infrastructure we
leveraged on the most recent advances in host interface technology,
physical link speed and connector mechanics; the result is the latest
generation of our hardware, the \apelinkp card --- see
Table~\ref{tab:card_features}. ---

The \apelinkp card is a single FPGA-based PCI Express board; the
employed FPGA device is the EP4SGX290, which is part of the
Altera\textsuperscript{\textregistered} 40 nm Stratix IV device family
and comes equipped with 36 full-duplex CDR-based transceivers,
supporting data rates up to 8.5 Gbps each. It also provides a PCIe X8
gen2 interface, which is complemented by a commercial PCIe core to
allow communication between the host processor and the network.

Moreover, an Ethernet port is foreseen in order to build an
additional, secondary network with an offload engine for collective
communication tasks.

\subsection{\apelinkp physical links}
In the global network structure, each card stands as a vertex of a 3D
torus mesh network with 6 independent point-to-point multiple links
channel (\ie the links between mesh sites).  Each link is made up of 4
bi-directional lanes bonded together; the automatic alignment logic is
our original addition.

Four links out of six are hosted on the main board; two more, say $Z+$
and $Z-$, are located in a detachable, small daughter-card on the
upper level.
In this way, the complete card takes on two PCI standard slots in a PC
chassis, mantaining the chance, if four links are enough, to use it in
a single slot wide configuration.
%
%

%
%

The torus links are 6 independent blocks with 2 virtual channel
receive buffers each, added to manage deadlock prevention.
Proper flow control is maintained via credits handshake between a
local RX block and the remote TX block; this handshake is embedded in
the link protocol data layer.
The torus link is able to autonomously re-transmit the header and the
footer in case of transmission errors.
Therefore, the protocol assures the delivery of the packet, avoiding
nonrecoverable situations where badly corrupted packets (with errors
in the header or footer) pose a threat to the global routing.
Packets with payload errors (signaled by the footer) must be instead
handled at the software level.
The chosen CRC polynomial generator is the industry-standard,
well-known CRC-32.

\subsection{\apelinkp routing capabilities}
The router comprises a fully connected, 7-ports-in/7-ports-out switch,
plus routing and arbitration blocks.
The routing block examines the header of each packet and translates
its destination address to a proper path across the switch; the
routing is dimension-ordered, with a measured latency of 60 ns.

\begin{table}[h]
\caption{\label{tab:card_features}Evolution of the \apelink cards.}
\begin{center}
\begin{tabular}{llll}
\br
 		&\bf \apelink		&\bf \apelinkp\\
\mr
\textbf{FPGA component}  & Altera Stratix S30 	& Altera Stratix IV GX 290 \\
\textbf{\# links}        & 6          		& 4/6\\
\textbf{link technology} & external National ser/des & embedded Altera transceivers  \\
\textbf{link cables}     & LVDS       & QSFP+ standard  \\
\textbf{raw link speed} & 6 Gbps     & 34 Gbps         \\
\textbf{host interface}  & PCI-X 133 MHz  & PCIe X8 Gen2  \\
\textbf{peak host BW}    & 1GB/s       & 4+4GB/s         \\
\br
\end{tabular}
\end{center}
\end{table}

%
\subsection{Test bed and preliminary results}
\label{apelink_test}
\begin{figure}[h]
\begin{minipage}{12pc}
\includegraphics[width=12pc]{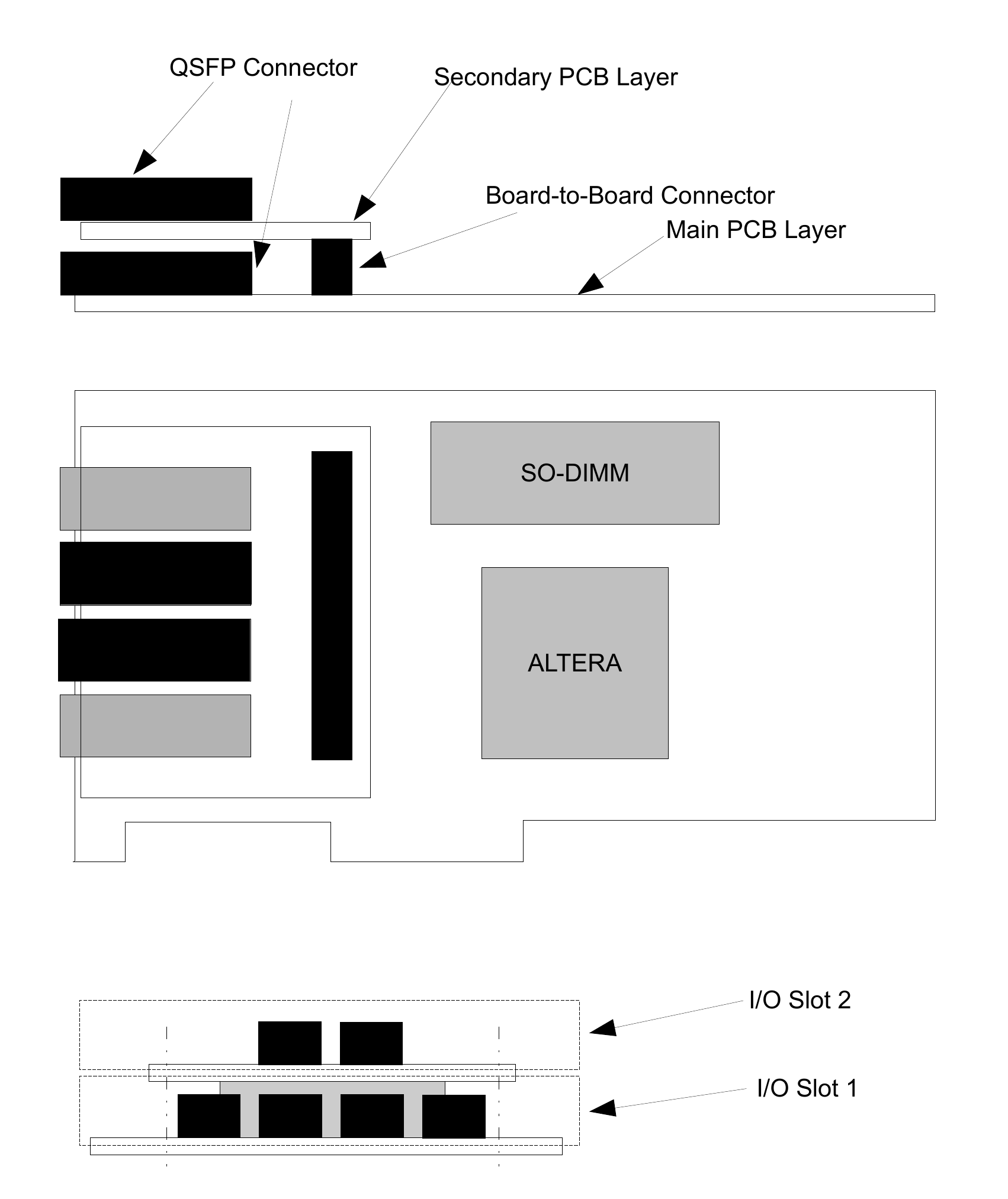}
\caption{\label{fig:board}\apelinkp board.}
\end{minipage}\hspace{2pc}%
\begin{minipage}{12pc}
\includegraphics[width=12pc]{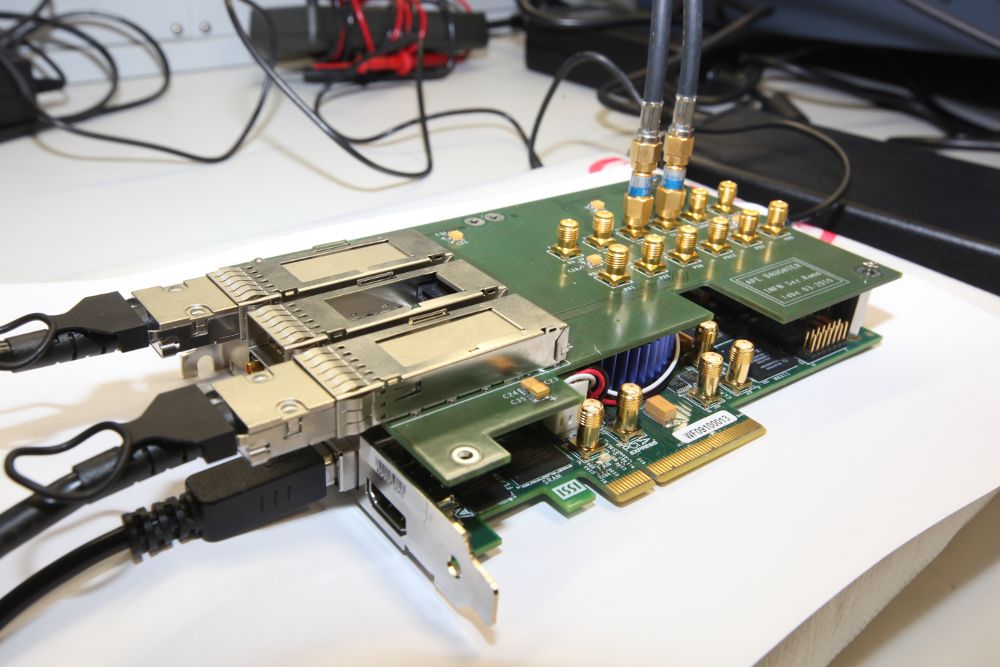}
\caption{\label{fig:test_system}Assembled \apelinkp test system.}
\end{minipage} 
\end{figure}

A schematic view of the complete \apenetp board is visible in
Figure~\ref{fig:board}.
The prototypes will be available at February 2011.


\begin{figure}[h]
\begin{minipage}{12pc}
\includegraphics[width=12pc]{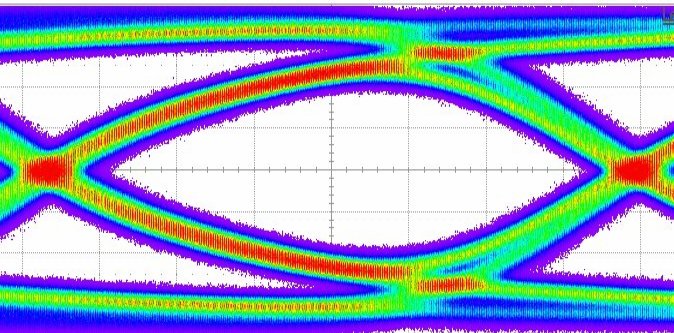}
\caption{\label{fig:5gbps_diretto}Eye diagram at 5~Gbps on the development kit.}
\end{minipage}\hspace{2pc}%
\begin{minipage}{12pc}
\includegraphics[width=12pc]{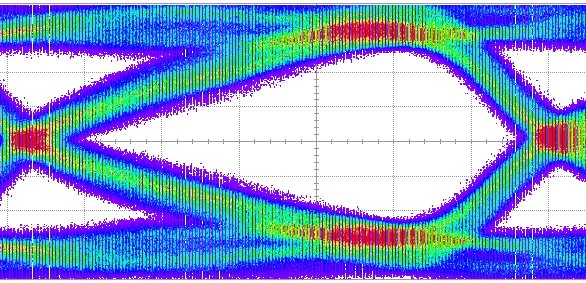}
\caption{\label{fig2b}Eye diagram at 3~Gbps on the mezzanine card.}
\end{minipage} 
\end{figure}

A test system has been built in order to develop the FPGA firmware,
the PCIe interface and the physical layer interconnection
technology~\cite{JINST}. We used a commercial
Altera\textsuperscript{\textregistered} development kit (equipped with
a smaller Altera\textsuperscript{\textregistered} Stratix IV GX 230)
and a custom-designed daughter-card (an HSMC mezzanine designed at
LABE in INFN-Roma) hosting 3 QSFP+ connectors and some SMA test
points.
This setup allows us to test the complete communication chain up to a
bitrate of 24 Gbps for each link.  Signal integrity was checked
connecting to dedicated SMA test points, straight at the output of the
FPGA transceivers (see Figure \ref{fig:5gbps_diretto}) and on the
mezzanine card (see Figure \ref{fig2b}) after one 
Samtec\textsuperscript{\textregistered} connector,
two QSFP+ connectors and 1 m of copper QSFP+ cable.

The link was successfully tested up to 3 Gbps data rate (compared to
8.5 Gbps achievable with the Stratix IV embedded transceivers).  Above
this limit we found some signal degradation probably caused by the
tower connector between the Altera development kit and the test
mezzanine. Investigation is in progress; a likely culprit is the
reduced bandwidth (below 5 GHz) of the 19 mm QTH 
Samtec\textsuperscript{\textregistered} connector,
which would be substituted anyway by higher bandwidth connectors in
the production release of the communication card ~\cite{samtec}.

Characterization of signal integrity (and maximum achievable
bandwidth) versus serial trans\-cei\-vers pre-emphasis and
equalization is still in progress.

\begin{figure}[h]
\begin{minipage}{12pc}
\includegraphics[width=12pc]{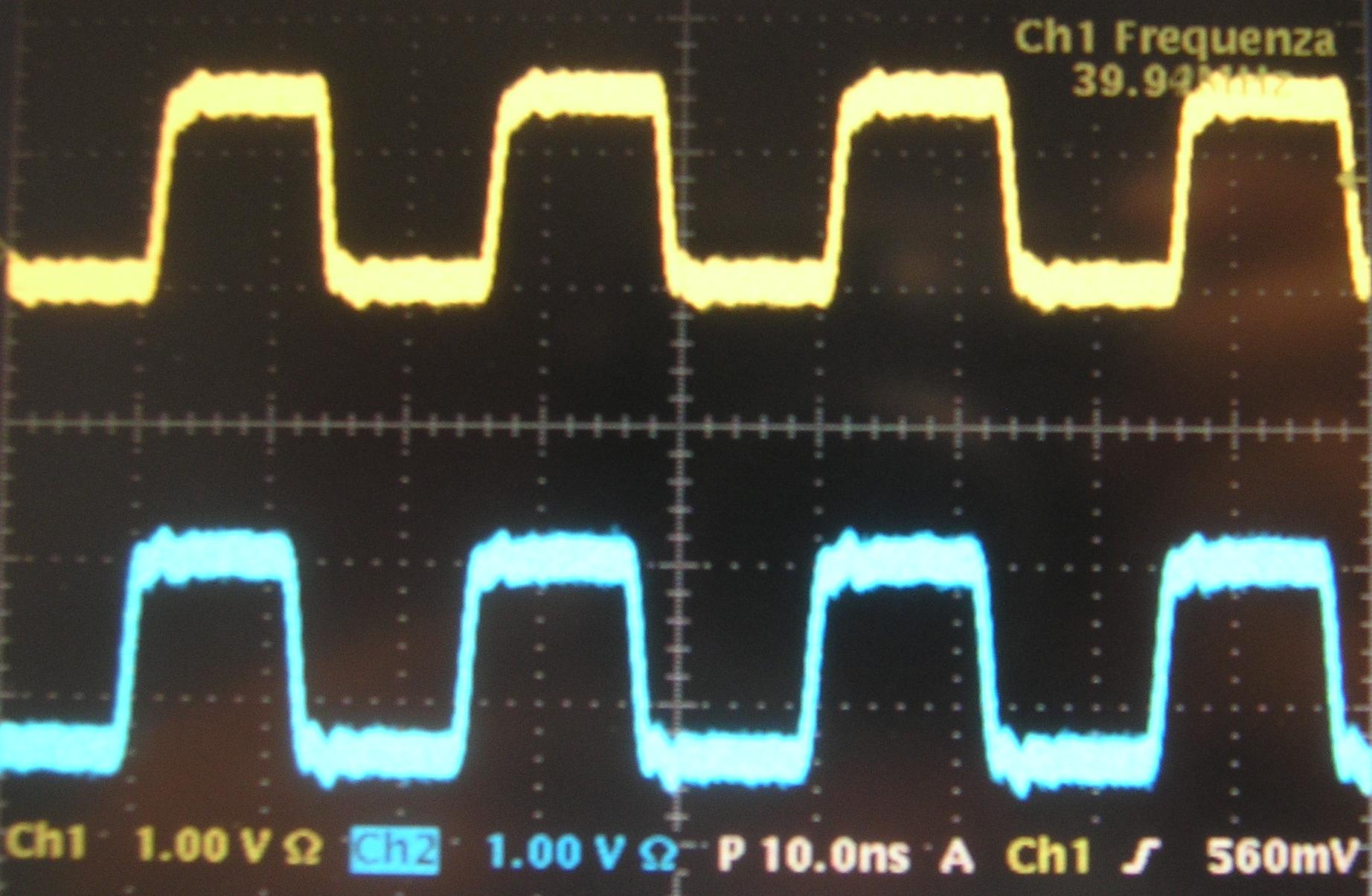}
\caption{\label{fig3a}Recovered clock stability check at 40 MHz.}
\end{minipage}\hspace{2pc}%
\begin{minipage}{12pc}
\includegraphics[width=12pc]{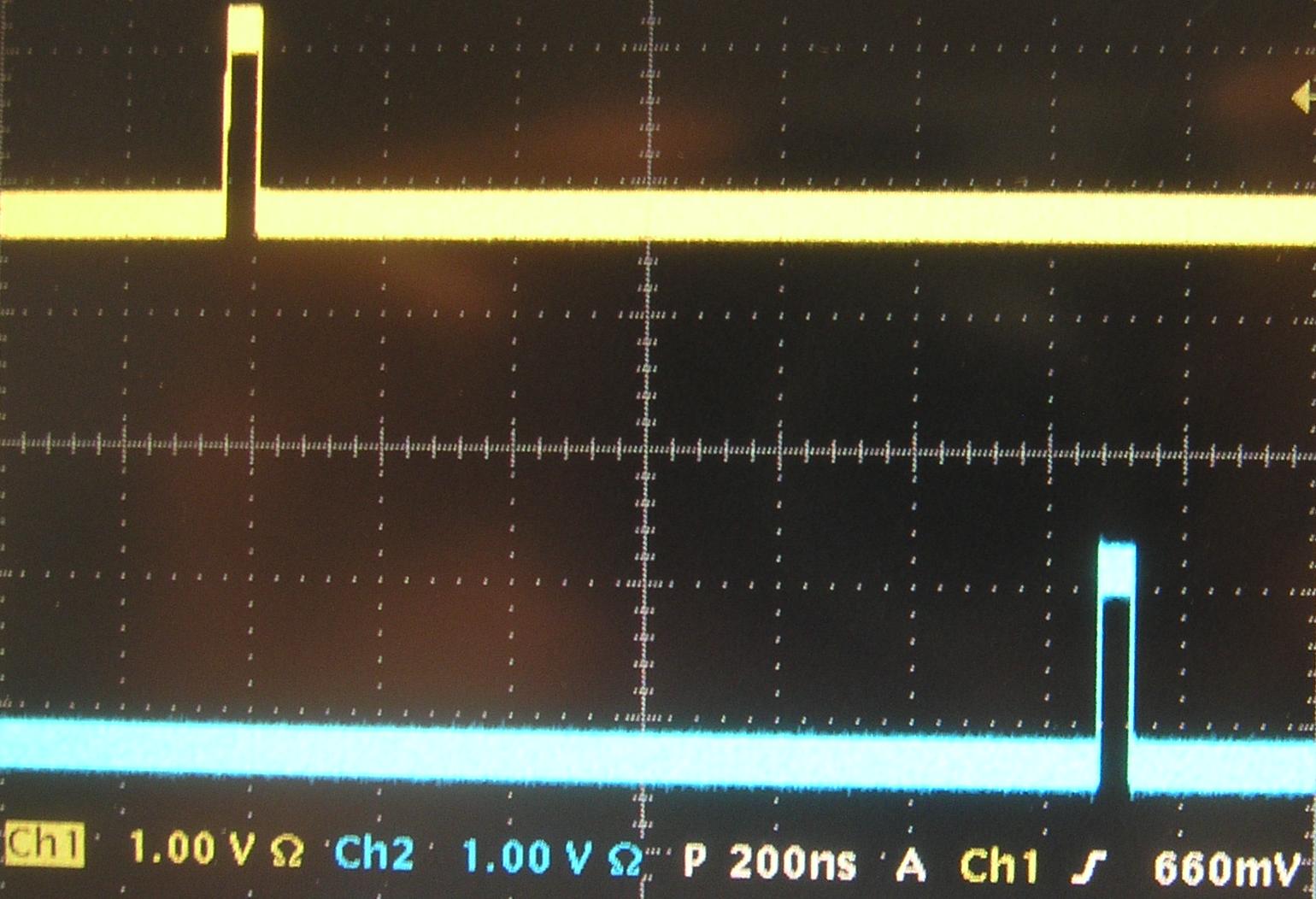}
\caption{\label{fig3b}Latency measurement at 40 MHz.}
\end{minipage} 
\end{figure}

Recovered clock stability was checked transmitting a pseudorandom data
stream organized in 128 bit wide words over 1 m copper QSFP+ cable and
checking the relative phase between the input and the output clocks
(see Figure \ref{fig3a}).  Recovered clock was found stable and in
phase with the input clock up to 400 MHz.

Latency was checked transmitting a pseudorandom sequence over 1 m
QSFP+ copper cable and rising a flag every time a fixed test word is
transmitted and received by the serializer and the deserializer
respectively (see Figure \ref{fig3b}). Transmission system latency was
found stable up to 160 MHz transmitting clock.

\section{The \apenetp software stack}
\label{sec:sw}
All \apenetp software is available under the GNU GPL Licence and spans
across four major topics: the firmware software running on the FPGA
embedded processor, the Linux kernel driver, the application level
RDMA library and a MPI implementation, these latter three developed and
tested under RedHat Enterprise Linux 5.
%
%

Communication primitives (\texttt{rmda\_put()}, \texttt{rdma\_get()},
\texttt{rdma\_send()}), buffer registration primitives
(\texttt{register\_buffer()}, \texttt{unregister\_buffer()}) and
synchronization primitives (\texttt{wait\_event()}) covering a custom
subset of the low-level RDMA APIs are made available to the
application programmer as a highly optimized C language library. On
top of these, we built a native \apenetp BTL module for OpenMPI 1.X.

Work is underway~\cite{gtc2010} on the hardware and software features
needed for GPU-initiated communications, \eg providing, using so
called PCIe peer-to-peer transactions, a CUDA-enabled ~\cite{cuda}
version of the \texttt{rdma\_put()} primitive, in order to avoid
intermediate copies onto CPU memory buffers.
To further reduce overhead, another development oversees the delivery
of RDMA events by the \apelinkp hardware in CPU memory in a way that
is accessible from within CUDA kernels.

Another research topic is exposing GPU memory areas as RDMA buffers,
in such a way they can be target of RDMA PUT and GET operations, even
more reducing the latency of network operations. To this end,
discussions are ongoing with some GPU vendors.

The firmware software running on the FPGA embedded processor is
currently in charge of managing the RDMA virtual-to-physical address
translation table, but we are exploring new ways to exploit it for
higher-level tasks.


\section{The deployment initiative}
\label{sec:quong}
%
We are currently exploring interconnection of GPU-equipped systems by
means of \apenetp (\quong project) to reach the PetaFLOPs range in
aggregated computing power and working on some GPU-related driver
optimizations.  For the 2011, our road-map foresees the integration of
a “\quong rack”, a mesh of computing nodes which are rack-mounted 1U
systems --– based on a commodity Intel CPU Xeon 5650 --– accelerated
via high-end GPUs (Nvidia Tesla C1060/M2050) interconnected with the
\apenetp hardware. This system, housed in a single rack of 42U, will
show a peak performance exceeding 60 TeraFLOPs and a power consumption
of less than 26~KW. Leveraging on \apenetp network, multiple \quong
racks can be assembled to push up the complete system to PetaFLOPs
scale.


\section{Conclusions and future developments}
\label{sec:devel}
A first mini-cluster is being assembled together with GPUs and the
\apelinkp version with 3 links, for final validation of the firmware,
the interconnection and the complete software stack on a small size
network (2-8 nodes).
Synthetic tests, as well as real life simulations, will be performed,
so to be ready with the 6-links prototype release and eventually a
bigger cluster deployment.

The presence on the APEnet+ card of a programmable component of
considerable power will allow us to explore reconfigurable computing,
\eg accelerating some tasks directly in hardware.

The needs of a large scale deployment make it necessary for \apenetp
to employ fault-tolerance features; we will be adding support for links'
self-diagnosis and the capability of routing around faulty
nodes~\cite{fault_tol, fault_tol2}.

\section{Acknowledgments}

The authors would like to thank the Electronics Laboratory at INFN
Sezione di Roma \cite{labe} for technical support with the design,
production and assembly of the test board used in this work.

This work was partially supported by the EU Framework Programme 7
project EURETILE under grant number 247846.

\section*{References}       

\smallskip


\end{document}